\documentclass[11pt,a4paper]{article}
\pdfoutput=1
\usepackage{jheppub}
\usepackage[utf8]{inputenc}
\usepackage[T1]{fontenc}
\usepackage[british]{babel}
\usepackage{lmodern}
\usepackage{amsmath}
\usepackage{amssymb}
\usepackage{amsfonts}
\usepackage{epsfig}
\usepackage{epstopdf}
\usepackage{graphicx,psfrag}
\usepackage{subcaption}

\usepackage{multirow}
\usepackage{cellspace} %
\newcommand{\beq}{\begin{equation}}
\newcommand{\eeq}{\end{equation}}
\newcommand{\bea}{\begin{eqnarray}}
\newcommand{\eea}{\end{eqnarray}}
\newcommand{\nn}{\nonumber}

\newcommand{\fig}{Fig.~}

\newcommand{\tr}{{\rm Tr}}
\newcommand{\bx}{{\bf x}}
\newcommand{\by}{{\bf y}}

\def\lsi{\raise0.3ex\hbox{$<$\kern-0.75em\raise-1.1ex\hbox{$\sim$}}}
\def\gsi{\raise0.3ex\hbox{$>$\kern-0.75em\raise-1.1ex\hbox{$\sim$}}}

\title{QCD in the heavy dense regime for general $N_c$\;:\\ On the existence of
quarkyonic matter}
\author{Owe Philipsen, Jonas Scheunert}
\affiliation{Institut f\"ur Theoretische Physik,
Goethe-Universit\"at Frankfurt am Main, \\ Max-von-Laue-Str. 1, 60438 Frankfurt am Main, Germany}

\emailAdd{philipsen, scheunert@itp.uni-frankfurt.de}

\abstract{
Lattice QCD with heavy quarks reduces to a three-dimensional effective 
theory of Polyakov loops, which is amenable to series expansion methods. 
We analyse the effective theory in the cold and dense regime for a general number of
colours, $N_c$. In particular, we investigate the transition from a hadron gas to 
baryon condensation. For any finite lattice spacing, we
find the transition to become stronger, i.e.~ultimately first-order, as $N_c$ is made large. 
Moreover, in the baryon condensed regime, we find the pressure to scale as $p\sim N_c$
through three orders in the hopping expansion. 
Such a phase differs from a hadron gas with $p\sim N_c^0$, or a quark gluon plasma, 
$p\sim N_c^2$, 
and was termed quarkyonic in the literature, since it shows both baryon-like and quark-like aspects.
A lattice filling with baryon number shows
a rapid and smooth transition from condensing baryons to a crystal
of saturated quark matter, due to the Pauli principle, and is consistent 
with this picture.  
For continuum physics,
the continuum limit needs to be taken before the large $N_c$ limit, which
is not yet possible in practice. However, in the controlled 
range of lattice spacings and $N_c$-values, 
our results are stable when 
the limits are approached in this order.
We discuss possible implications for physical QCD.
}

\begin{document}
\maketitle


\section{Introduction}

The QCD phase diagram is important for many aspects
of current nuclear, heavy ion and astro-particle physics, yet it remains
largely unknown. This is
because lattice QCD at finite baryon chemical potential has a severe sign problem,
which prohibits straightforward Monte Carlo simulations. Various workarounds
to extend the Monte Carlo method introduce
additional approximations and are limited to the high temperature and/or
low density region, with baryon chemical potential $\mu_B/T\lsi 3$ \cite{review}.
No sign of criticality is found in this region, where the transition from a hadronic 
gas to a quark gluon plasma proceeds by an analytic crossover \cite{aoki,pastor,bazavov, vov}.
The same conclusion is reached by analytic non-perturbative approaches like
Dyson-Schwinger equations \cite{cf} or the functional renormalisation group \cite{jp}.
Despite continuing efforts, a genuine solution to the sign problem, and hence fully 
non-perturbative access to  
the cold and dense region of QCD, are missing to date.

This situation has motivated the study of QCD
also in unphysical, but controllable parameter regions, where the sign problem can be 
overcome by either algorithmic or analytic methods. 
In this work our goal is to bridge two such approaches. 
We employ an effective lattice theory derived from the standard Wilson action 
by combined strong coupling and hopping (inverse quark mass) expansions. 
The effective theory is valid on reasonably fine lattices,
as long as the quarks are sufficiently heavy 
to be described by the next-to-next-to-leading order in the hopping expansion.
In this parameter range the theory correctly reproduces the critical heavy quark mass at 
zero density, 
where the first-order deconfinement transition changes to a smooth 
crossover \cite{fromm}, 
and furthermore allows for an extension to finite baryon chemical potential, including
the cold and dense regime around the onset transition to baryon matter. 

Here we consider the effective theory for general colour 
gauge group $SU(N_c)$, in order to establish 
contact with another effective approach in the continuum, namely QCD at large $N_c$. 
In particular, we analyse thermodynamic functions around 
the onset transition to baryon matter in the cold and dense regime,
for varying and large $N_c$.
This allows us to address, by direct calculation, various conjectures made in \cite{quarky}  
regarding the phase diagram and the effective degrees of freedom at large $N_c$.
There, the authors argue for the existence of quarkyonic matter, which is characterised by its pressure
scaling as $p\sim N_c$ and has both baryon-like and quark-like aspects. Phenomenological 
consequences of this form of matter in physical QCD have been assessed in \cite{anton,spirals,pheno,reddy}.
A general, qualitative discussion about the 
possibilities for the phase diagram in $(T,\mu,N_c)$-space as well as references to earlier work can be found in \cite{mish}.

We begin with a brief review of the effective lattice theory in 
section \ref{sec:efft}.
This material is not new and can be skipped by readers familiar with it, 
but is needed as reference point when interpreting the
following results. In section \ref{sec:nc}, a summary of the conjectured 
phase diagram at large $N_c$ 
is followed by our proper calculations for general $N_c$ and the analysis 
of the results for large $N_c$.
Finally, section \ref{sec:phys} concludes what is expected for
physical QCD.
 
\section{QCD with heavy quarks  \label{sec:efft}}

\subsection{The effective lattice theory}

Consider the partition function of lattice QCD with the standard Wilson action at 
finite temperature,  $T=1/(aN_\tau)$, realised by compact euclidean time 
with $N_\tau$ slices and (anti-) periodic boundary conditions for (fermions) bosons. 
An entirely equivalent formulation in terms of temporal lattice links only
is obtained after performing the Gauss integral over the quark fields and integrating the gauge links in spatial directions,
\beq
Z=\int DU_0DU_i\;\det Q \; e^{-S_g[U]}\equiv\int DU_0\;e^{-S_{\mathrm{eff}}[U_0]}=\int DW \,\;e^{-S_{\mathrm{eff}}[W]}\;.
\eeq 
With the spatial links gone, the effective action depends on the temporal links 
only via Wilson lines closing through 
the periodic boundary, 
\beq
W({\bf x})=\prod_{\tau=1}^{N_\tau} U_0({\bf x},\tau)\;. 
\eeq
For $SU(2)$ and $SU(3)$ the effective action can always be expressed in terms of 
Polyakov loops, $L({\bf x})=\tr W({\bf x})$, whereas for larger $N_c$ in general
traces of higher powers of $W$ appear as well.

This effective action is unique and exact. However, 
the integration over spatial links causes long-range interactions 
of Polyakov loops at all distances and to all powers so that in practice truncations are necessary. 
For non-perturbative ways to define and determine truncated 
theories, see \cite{wozar,green1,green2,bergner}.
Here, we use an effective theory based on expanding the path integral in a combined character and
hopping parameter series.
Both expansions result in convergent series within a finite radius of
convergence (for an introduction, see \cite{mm}). Truncating these at some finite order, the integration over
the spatial gauge links can be performed analytically to provide a closed
expression for the effective theory. Going via an effective action results in a
resummation to all powers with better convergence properties compared to a
direct series expansion of thermodynamic observables as in \cite{lange1,lange2,p_nc}.
Since the Wilson line $W({\bf x})$ contains the length $N_\tau$ of the
temporal lattice extent implicitly, the effective theory is three-dimensional.
Note that, in the case of 4d Yang-Mills theory, this representation by a 3d
centre-symmetric effective theory is the basis for the Svetitsky-Yaffe
conjecture \cite{sy,polon} concerning the universality of $SU(N_c)$ deconfinement transitions.
Including the quark determinant via the hopping expansion introduces centre
symmetry breaking terms and additional effective couplings.

Let us briefly summarise the expansions used in order to perform the
spatial link integrations.
The gauge part of the action is a class function with respect to the product of
the links of one plaquette:
\begin{equation}
  S_g[U] = \sum_{p} S_{g,p}(U_p) = \sum_{p} S_{g,p}(V^{-1} U_p V),
\end{equation}
where $V \in SU(N_c)$.
Therefore it can be expanded in the characters $\chi_r$ of the irreducible
representations $r$ of $SU(N_c)$ at every plaquette,
\begin{equation}
  \label{eq:character-expansion} 
	e^{-S_{g,p}(U_p)} = c_0\Big(1 + \sum_{r\neq 0} d_r a_r(\beta) \chi_r(U_p)\Big)\;.
\end{equation}
In this formula, $d_r$ denotes the dimension of the representation
and $a_r(\beta)$ is the character expansion coefficient divided
by the expansion coefficient of the trivial representation. The expansion
coefficients can be computed exploiting the orthogonality of the characters:
\begin{align}
  a_r(\beta) & = \frac{c_r(\beta)}{c_0(\beta)}, \\
  c_r(\beta) & = \int\limits_{SU(N_c)} dU \chi_r(U)^\ast \exp(-S_{g,p}(U)).
\end{align}
We drop the overall factor of $c_0$ in
equation~\eqref{eq:character-expansion}, as it cancels in expectation values.
For the integration of the spatial links following this expansion, 
one can use the formulas
\begin{align}
  \label{eq:character-integration-1}
  \int\limits_{SU(N_c)} dU \; \chi_r(U V) \chi_s(W U^{-1})
  & = 
  \delta_{r s} \frac{1}{d_r} \chi_r(V W)\;, \\
  \label{eq:character-integration-2}
  \int\limits_{SU(N_c)} dU \; \chi_r(U V U^{-1} W)
  & = 
  \frac{1}{d_r} \chi_r(V) \chi_r(W)\;, 
\end{align}
for those cases where not more than two non-trivial representations share a
common link. In earlier publications \cite{efft1} this was used to derive 
the effective gauge
action for $SU(3)$ to rather high orders in the coefficient of the 
fundamental character $u(\beta)\equiv a_f(\beta)/d_f = a_f(\beta)/N_c$.
The coefficients of the higher dimensional representations can be
expressed in terms of the fundamental one, see also section
\ref{sec:efft-general-nc}, and therefore the expansion can be organised according to
powers of the fundamental character. 
The dependence of $u$ on the lattice gauge coupling
$\beta=2N_c/g^2$ can be specified either as a power series or numerically, 
\begin{equation}
  u(\beta)=\frac{\beta}{18}+\frac{\beta^2}{216}+\ldots < 1.
\end{equation}
It is 
known to arbitrary precision, and $u$ is always smaller than
one for finite $\beta$-values.

For the hopping expansion it is useful to split the quark matrix in
temporal and spatial hops between nearest neighbours,
\begin{align}
  Q
  & = 1 - T - S, \\
  \det(Q)
  & = \det(1-T) \det(1-(1-T)^{-1} S) \\
  & = \det(Q_{\mathrm{stat}}) \det({Q_{\mathrm{kin}}})\;.
\end{align}
This is because the static determinant containing all temporal hops (and only those) can be computed exactly \cite{hetrick, bind}.
We then do the hopping expansion of the kinetic quark determinant using
\begin{equation}
  \det{Q_{\mathrm{kin}}} = \exp(\tr(\ln(Q_{\mathrm{kin}})))\;.
\end{equation}
This leads to an expansion in powers of $S$, which is proportional to the hopping parameter, 
\begin{equation}
  \kappa=\frac{1}{2am_q+8}\;.
\end{equation}
The expansion terms are then ordered according to their number of spatial hops
while the temporal ones are resummed to all orders. Since the hopping expansion is
in inverse quark mass, the effective theory to low orders is valid for heavy
quarks only. For a derivation of the effective theory to order
$\mathcal{O}(\kappa^4)$ in spatial hops, see \cite{bind}. In this case the
relevant integrals for the fermionic contributions are
\begin{align}
  \label{eq:fermion-integrals-1} 
  \int\limits_{SU(N_c)} dU \; U_{ij} U_{kl}^\dagger 
  & = \frac{1}{N_c} \delta_{il} \delta_{jk}, \\
  \label{eq:fermion-integrals-2}
  \begin{split}
    \int\limits_{SU(N_c)} dU
    \;U_{i_1j_1}U_{i_2j_2}U^\dagger_{k_1l_1}U^\dagger_{k_2l_2} & =
    \frac{1}{N_c^2-1}
    \Big[\delta_{i_1l_1}\delta_{i_2l_2}\delta_{j_1k_1}\delta_{j_2k_2} +
    \delta_{i_1l_2}\delta_{i_2l_1}\delta_{j_1k_2}\delta_{j_2k_1}\Big] \\
    & \quad -\frac{1}{N_c(N_c^2-1)}
    \Big[\delta_{i_1l_2}\delta_{i_2l_1}\delta_{j_1k_1}\delta_{j_2k_2} +
    \delta_{i_1l_1}\delta_{i_2l_2}\delta_{j_1k_2}\delta_{j_2k_1}\Big].
  \end{split}
\end{align}

Generically, the effective action obtained in this way has the
following form:
\begin{eqnarray}
-S_{\mathrm{eff}}=\sum_{i=1}^\infty\lambda_i(u,\kappa,N_\tau)S_i^s-
2N_f\sum_{i=1}^\infty\left[h_i(u,\kappa,\mu,N_\tau)S_i^a+
\bar{h}_i(u,\kappa,\mu,N_\tau)S_i^{a,\dagger}\right]\;.
\label{eq_defseff}
\end{eqnarray}
The $\lambda_i$ are defined as the effective couplings of the 
$Z(N_c)$-symmetric terms $S_i^s$, whereas the $h_i$ multiply the asymmetric 
terms $S_i^a$. 
In particular, $h_1, \bar{h}_1$ are the coefficients of $L,L^*$, respectively, and
to leading order correspond to the fugacity of the quarks and anti-quarks,
\bea
\label{eq:h1}
h_1&=&(2\kappa)^{N_\tau}e^{a\mu N_\tau}(1+\ldots) =h_1^{\rm LO}(1+\ldots)
=e^{\frac{\mu-m}{T}}(1+\ldots)\;,\\  
\bar{h}_1&=&(2\kappa)^{N_\tau}e^{-a\mu}(1+\ldots) =\bar{h}_1^{\rm LO}(1+\ldots)
=e^{-\frac{\mu+m}{T}}(1+\ldots)\;.
\eea
Here, 
\beq
am=\ln(2\kappa)=\frac{am_B^{\mathrm{LO}}}{N_c}
\label{eq:mlo}
\eeq
is the constituent quark mass in lattice units 
of a baryon computed to leading order in the hopping expansion \cite{ks}, 
while 
\beq
h_2=\kappa^2 N_\tau/N_c(1+\ldots)
\eeq
is the effective coupling of a nearest neighbour $L_\bx L_\by$ interaction.

The partition function for $SU(3)$, including just these simplest interactions, reads
\bea
\label{zpt}
Z&=&\int DW\prod_{<\bx, \by>}\left[1+\lambda_1(L_{\bx}L_{\by}^*+L_{\bx}^*L_{\by})\right]\\
&&\times\prod_{\bx}[1+h_1L_{\bx}+h_1^2L_{\bx}^*+h_1^3]^{2N_f}[1+\bar{h}_1L^*_{\bx}+\bar{h}_1^2L_{\bx}+\bar{h}_1^3]^{2N_f}
\nonumber\\
&&\times \prod_{<\bx, \by>}\left(1-h_{2}{\rm Tr} \frac{h_1W_{\bx}}{1+h_1W_{\bx}}{\rm Tr} \frac{h_1W_{\by}}{1+h_1W_{\by}}\right)
\left(1-h_{2}{\rm Tr} \frac{\bar{h}_1W^\dag_{\bx}}{1+\bar{h_1}W^\dag_{\bx}}{\rm Tr} \frac{\bar{h}_1W^\dag_{\by}}{1+\bar{h}_1W^\dag_{\by}}\right)
\times \ldots \;.\nn
\eea
In this expression the first line represents the pure gauge sector, the second line is the static determinant and the third line
the leading correction from spatial quark hops. This partition function has a 
weak sign problem and can be simulated with 
either reweighting or complex Langevin methods \cite{fromm,bind}.
Since the effective couplings correspond to power series of the expansion parameters, 
they are themselves small in the range of validity. 
Hence, the effective theory can also be treated by linked-cluster expansion methods 
known from statistical physics,
with results for thermodynamic observables in quantitative agreement with the 
numerical ones \cite{k8}. 
In this way, full control over the sign problem is achieved.

\subsection{The deconfinement transition} 

\begin{figure}
\centerline{
\includegraphics[width=0.5\textwidth]{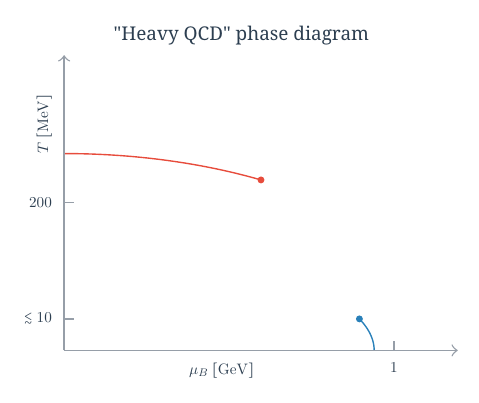}
}
\caption[]{The phase diagram of QCD with very heavy quarks.}
\label{fig:pd_heavy}
\end{figure}
The phase diagram of QCD with heavy quarks is depicted schematically in \fig\ref{fig:pd_heavy}. 
At zero density, the thermal transition is a first-order deconfinement transition. It is
a remnant of the centre symmetry-breaking transition of the $SU(3)$ pure gauge theory, which
gets weakened by explicitly centre-breaking finite quark masses $\sim 1/m_q$, until
it ends in a second-order point for some critical mass $m_q^c$.
In the effective theory, this phase transition appears as spontaneous breaking of the 
$Z(3)$-symmetry 
at some set of critical couplings 
$\lambda_{i,c}=\lambda_i(u_c, \kappa_c, N_\tau), h_{i,c}(u_c,\kappa_c,N_\tau)$, which can 
be determined by numerical simulation. Inversion of the
effective couplings then gives predictions for 
$\beta_c(N_\tau), \kappa_c(N_\tau)$, which can be compared with the results from full QCD simulations.

For $SU(2)$ and $SU(3)$-Yang-Mills theory, the simplest effective theory 
with only a nearest neighbour coupling 
(first line in equation (\ref{zpt}))
correctly reproduces the universality of the respective 
deconfinement transitions,
and the predicted $\beta_c(N_\tau)$ are within 10\% of 
their true values for $N_\tau=2,\ldots,16$ \cite{efft1}.
For QCD with heavy quarks, 
the simplest effective theory including the static determinant and 
$\kappa^2$-corrections,
predicts $\kappa_c$ to better than 10\% on $N_\tau=4$ \cite{fromm}.
Contrary to full QCD, the effective theory can be simulated at finite 
chemical potential to determine the location of the
critical end point as a function of quark mass \cite{fromm}.
This qualitative behaviour of the deconfinement transition in the heavy quark
region is also found 
by continuum studies using a Polyakov loop model \cite{lo} and in the
functional renormalisation group approach \cite{frg}.

\subsection{The onset transition to finite baryon number}
\label{sec:review}

Going out along the chemical potential axis at low temperature, the system crosses the onset transition beyond which
the ground state consists of condensed baryon matter. 
In order to interpret our analysis for general $N_c$, let us first 
recall the situation for $N_c=3$ in some detail.
The qualitative features are best understood
in the strong coupling limit, $\beta=0$, with the static quark determinant only, where (\ref{zpt}) is reduced to the second line.  
The partition function then
factorises into one-site integrals which can be solved analytically. 
Since we are interested in low temperatures, where
mesonic contributions are exponentially suppressed by their fugacity factors, 
we simplify the analysis by setting $\bar{h_1}=0$.
For $N_f=1$ the partition function then reads  \cite{silver, bind} 
\beq
\label{eq:z-0-su3}
Z(\beta=0) \stackrel{T\rightarrow 0}{\longrightarrow}z_0^V \quad \mbox{with}\quad 
z_0=1+4h_1^{3}+h_1^{6}\;,
\eeq
corresponding to a free baryon gas with two species. 
With one quark flavour only, there are no nucleons and the
first prefactor indicates a spin 3/2  quadruplet of $\Delta$-baryons whereas 
the second term is a spin 0 six-quark state or di-baryon.
The quark number density is 
\beq
n=
\frac{T}{V}\frac{\partial}{\partial \mu}\ln Z=\frac{1}{a^3}\frac{4N_ch_1^{N_c}+2N_ch_1^{2N_c}}{1+4h_1^{N_c}+h_1^{2N_c}}\;,
 \quad \lim_{T\rightarrow 0} a^3n=\left\{\begin{array}{cc} 0, & \mu<m\\
	2N_c, & \mu>m\end{array}\right.\;,
\eeq
and at zero temperature exhibits a discontinuity when the quark chemical potential equals the constituent mass $m$.
This reflects the ``silver blaze'' property of QCD, i.e.~the fact that the baryon number stays zero
for small $\mu$ even though the partition function explicitly depends on it \cite{cohen}. Once the baryon chemical potential 
$\mu_B=3 \mu$
is large enough to make a baryon ($m_B=3m=m_B^\mathrm{LO}$ in the static strong coupling limit), a discontinuous phase transition 
to a saturated crystal takes place. 
Note that saturation density here is $2N_c$ quarks per flavour and lattice
site and reflects the Pauli principle. This is clearly a discretisation effect that disappears
in the continuum limit.

For two flavours the corresponding expression for the free baryon gas reads
\begin{eqnarray}
z_0& =& (1 + 4 h_d^3 + h_d^6)+ (6 h_d^2 + 4 h_d^5) h_u+ (6 h_d + 10 h_d^4)h_u^2+ 
  (4 + 20 h_d^3 + 4 h_d^6)h_u^3 \nn \\
&&  + (10 h_d^2 + 6 h_d^5) h_u^4+ ( 4 h_d + 6 h_d^4) h_u^5 
  +(1 + 4 h_d^3 + h_d^6)h_u^6\;,
\label{eq:freegas}  
\end{eqnarray}
where we have now distinguished between the $h_1$ coupling for the $u$- and $d$-quarks. In this case we identify in addition the 
spin 1/2 nucleons as well as many other baryonic multi-quark states with their correct spin degeneracy. A similar result is obtained for
mesons if we instead consider an isospin chemical potential in the low temperature 
limit \cite{bind}. 
Again, the onset transition to finite baryon density is a step function from zero
to saturation density, which now is $2N_cN_f$ quarks per site.

The step function behaviour gets immediately smeared out to a smooth crossover, as soon as a finite temperature, $N_\tau<\infty$,
is switched on. This implies 
that the first-order line of the nuclear liquid gas transition is
exponentially short as a result of the large quark masses, as expected from nuclear physics Yukawa potentials with meson exchange. 
Indeed, the interaction energy per baryon, which sets the scale for the critical endpoint of the nuclear liquid gas transition, 
can be extracted from the 
dimensionless combination of energy density and baryon number of the system,
\beq
\epsilon(\mu,T)=\frac{e(\mu,T)-n_B(\mu,T) m_B}{n_B(\mu,T) m_B}\;,
\eeq  
in the limit of zero temperature. In the strong-coupling limit one finds 
\beq
\epsilon= -\frac{4}{3}\frac{1}{a^3n_B}\left(\frac{6h_1^3+3h_1^6}{z_0}\right)^2\,\kappa^2+\ldots\;.
\eeq
Thus the length of the liquid gas transition in \fig\ref{fig:pd_heavy} 
is a function of quark mass and decreases to zero 
towards the static limit. Including $\kappa^4$-corrections, a first-order transition
ending at some finite temperature is explicitly seen \cite{bind}.

Including the gauge coupling, and with sufficiently many corrections at hand, also the
lattice spacing can be varied and the approach to the continuum can be studied. In \cite{k8} it was shown that 
through orders $u^5\kappa^8$ for a sufficiently heavy quark mass, continuum-like behaviour is obtained immediately after
the onset transition, with the qualitative features discussed here. 

\section{QCD for large \texorpdfstring{$N_c$}{Nc} \label{sec:nc}}

Since in the framework of the effective theory we can work fully analytically, 
it is straightforward to investigate
what happens when the number of colours $N_c$ is varied and made large. 
In particular, we aim to explore some 
large $N_c$ considerations leading to  the prediction of quarkyonic matter \cite{quarky}.

There is a lot of interesting literature on QCD at large $N_c$, 
which we are unable to represent properly. In particular, 
baryon matter in the combined heavy quark and large $N_c$ limits 
has been considered by a mean field analysis in the continuum \cite{cohen2,cohen3}. 
Here, our approach is quite different in working on the lattice with large
but finite quark masses, for general $N_c$ and beyond mean field. 

The essential qualitative features of QCD at lage $N_c$ were 
established in the early works \cite{hooft,witten}.
The 't Hooft limit is defined by 
\beq
N_c\rightarrow \infty\quad \mbox{with} \quad \lambda_H\equiv g^2N_c={\rm const.}
\label{eq:limit}
\eeq
In this case the theory has the following properties:
\begin{itemize}
\item Quark loops in Feynman diagrams are suppressed by $N_c^{-1}$
\item Non-planar Feynman diagrams are suppressed by $N_c^{-2}$
\item Mesons are free; the leading corrections are cubic interactions $\sim N_c^{-1/2}$ and quartic interactions $\sim N_c^{-1}$
\item Meson masses are $\sim \Lambda_{QCD}$
\item Baryons consist of $N_c$ quarks, baryon masses are $\sim N_c \Lambda_{QCD}$
\item Baryon interactions are $\sim N_c$
\end{itemize}

\begin{figure}[t]
\centering
\includegraphics[width=7cm]{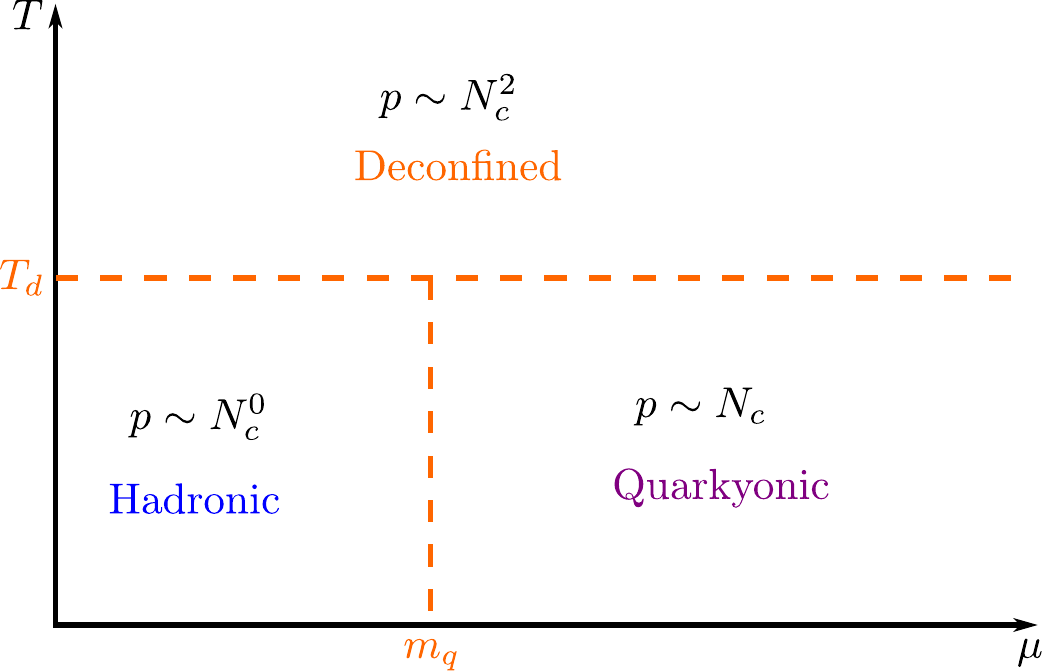}\hspace*{2cm}
\includegraphics[width=4cm]{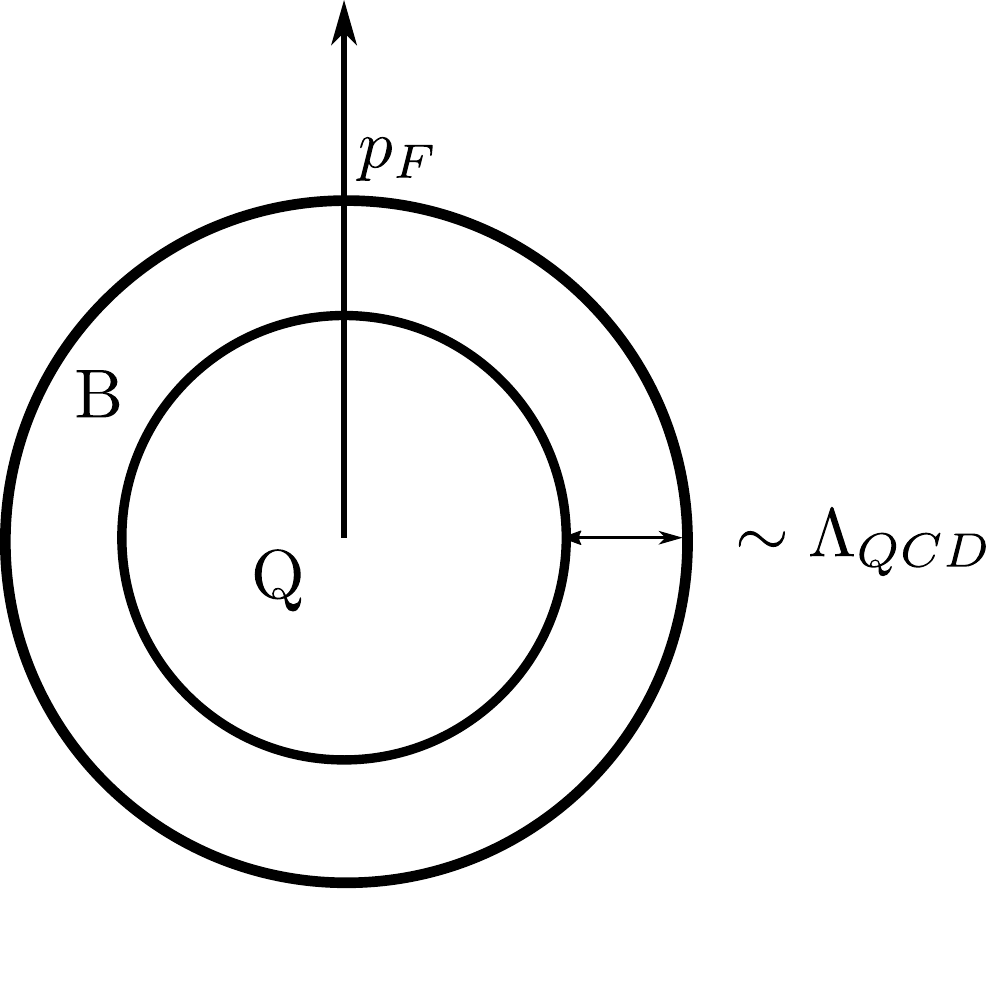}
\caption[]{Left: Phase diagram in the limit of large $N_c$, as conjectured in \cite{quarky}. 
Right: Quarkyonic matter in momentum space, with quark matter (Q)
surrounded by a shell of baryons (B).
}
\label{fig:pd_nc}       
\end{figure}

The authors of \cite{quarky} used these and various other ingredients to draw qualitative conclusions 
for the QCD phase diagram.
\fig\ref{fig:pd_nc} shows their conjectured phase diagram in the large $N_c$ limit.
From finite temperature perturbation theory it follows that, 
in the plasma phase, $p\sim N_c^2$. 
With quark loops suppressed, the phase boundary 
of the deconfinement transition is pure gauge-like and unaffected by 
chemical potential. 
It thus forms a horizontal line, 
staying first-order everywhere. 
On the other hand, in the hadronic, low density phase, 
thermodynamics is 
quantitatively well described by a weakly interacting hadron resonance 
gas \cite{wb,hotQCD}.
Statistical mechanics then implies that
the baryonic contribution to the pressure is exponentially suppressed 
with baryon mass, so $p\sim N_c^0$ there.  In \cite{quarky} a
similar combination of perturbative (valid for large $\mu$) and statistical mechanics arguments 
for baryons suggests that, for  
low temperatures and $\mu_B>m_B$, 
the pressure scales as $p\sim N_c$. The authors termed this phase ``quarkyonic'', since it 
shows aspects of both quark matter and baryon matter.
In particular it is argued that, for zero temperature, excitations relative to the Fermi
sea should be baryon-like for $(p-p_F)\lsi \Lambda_{QCD}$ and quark-like for 
$(p-p_F)\gg \Lambda_{QCD}$, implying a shell structure in momentum space as in 
\fig\ref{fig:pd_nc} (right). Since the Fermi momentum is $p_F\lsi \Lambda_{QCD}$ at the onset transition and 
then grows with quark chemical potential, this picture suggests the 
possibility to smoothly interpolate between baryon matter (right after the onset)  and quark matter (at
very large densities).  
We will now address these issues by direct calculations using
the effective lattice theory.

\subsection{The effective lattice theory for general \texorpdfstring{$N_c$}{Nc}} \label{sec:efft-general-nc}

Note that
$N_c=2$ has already been analysed in detail \cite{scior}, with interesting
physics results for two-colour QCD. Our aim here is to go in the other direction
and to increase $N_c$. 
For the gluonic part, the derivation of the effective theory for general and
large $N_c$ has already been discussed in \cite{myers}. Note that the integration rules~\eqref{eq:character-integration-1} and
\eqref{eq:character-integration-2} are true for arbitrary $N_c$. Therefore, in
cases where only these formulas are relevant, one simply has to replace $d_r$
and $a_r$ by their appropriate generalisations to $N_c$.
Specifically, the character expansion coefficients $a_r$ can be obtained via
\cite{Drouffe:1983fv}
\begin{equation}
  \label{eq:character-coeff-all-nc}
  a_r(\beta/2 N_c) = 
  \frac{1}{d_r} 
  \sum_{n=-\infty}^\infty \det_{1\leq i,j \leq \infty}
  \left(I_{\alpha_j+i-j+n}(\beta/2N_c)\right)
  \bigg/
  \sum_{n=-\infty}^\infty \det_{1\leq i,j \leq \infty}
  \left(I_{i-j+n}(\beta/2N_c)\right).
\end{equation}
In this formula, the $\alpha_i$ are a set of $N_c$ positive descending
integers with $\alpha_{N_c}=0$ which label the representation and correspond to
Young tableaux. Following \cite{Drouffe:1983fv} one may re-express all
coefficients of higher representations in terms of the fundamental
representation using double Young tableaux. The characters corresponding to a
double young tableau can be determined in terms of the traces of powers of $U$
and $U^{\dagger}$ \cite{Green:1980bg}.

In \cite{myers} the fermionic contributions were also discussed using the hopping expansion. 
However, only
a subset of spatial hoppings to
$\mathcal{O}(\kappa^4)$ was considered, and temporal hoppings were included up to
$\mathcal{O}(\kappa^{2N_\tau})$. As we mentioned in section \ref{sec:efft}, we work in a scheme
where temporal hoppings are resummed to all orders. Nevertheless,
equations \eqref{eq:fermion-integrals-1} and \eqref{eq:fermion-integrals-2}
are valid for general $N_c$ so the fermionic contributions obtained in this way
are legitimate also for general $N_c$. \\
Spatial baryon hoppings contribute at $\mathcal{O}(\kappa^{N_c})$ and
therefore they are suppressed for large $N_c$. To evaluate the contributions of
meson hoppings, integrals of the type 
\begin{equation} 
  \int dU \; U_{i_1 j_1}
  \cdots U_{i_a j_a} U_{k_1 l_1}^\dagger \cdots U_{k_a l_a}^\dagger
\end{equation} 
are needed. Since these integrals give the same result
for $U\in U(N_c)$ and $U\in SU(N_c)$ one can, for the fermionic
contributions in the vacuum and at finite temperature, work with $U(N_c)$ instead of $SU(N_c)$ at large $N_c$.
Likewise, when analysing the pure gauge theory for large $N_c$, the same
simplification applies \cite{gw}.  
However, at finite chemical potential
temporal quark hoppings in the positive direction are boosted by a factor of
$e^{a\mu}$ and therefore integrals of the type
\begin{equation}
  \int dU \; U_{i_1 j_1}
  \cdots U_{i_a j_a} U_{k_1 l_1}^\dagger \cdots U_{k_b l_b}^\dagger
\end{equation}
with
\begin{equation}
  b - a = 0 \text{ mod } N_c
\end{equation}
are relevant. These integrals vanish for $U(N_c)$
when $b \neq a$, but not for $SU(N_c)$, where they contain baryonic contributions. 
For the cold and dense regime, we therefore need to calculate integrals over $SU(N_c)$.

\subsection{Evaluation of the  \texorpdfstring{$SU(N_c)$}{SU(Nc)} effective theory in the strong coupling limit}

We begin our analysis in the strong coupling limit, $u(\beta=0)=0$ 
and thus $\lambda_1=0, h_1=h_1^{\rm LO}$, 
and employ the effective theory to order 
$\mathcal{O}(\kappa^4)$ in spatial hoppings. 
We also neglect terms containing $\bar{h}_1$, since they are exponentially suppressed
at low temperatures. 
The theory with these approximations already shows the most salient features of baryon dynamics, and we discuss
modifications by the neglected couplings in later sections. 
Thus we get the free energy density
\begin{align}
  \label{eq:free-energy}
  \begin{split}
    -f 
    &= 
    \ln(z_0) - 6 N_f \frac{\kappa^2 N_\tau}{N_c}
    \left(\frac{z_{(11)}}{z_0}\right)^2 \\
    &\phantom{{}={}}
    + 3 \frac{\kappa^4 N_{\tau}}{2(N_c^2-1)} 
    \frac{4 N_f^2 z_{(22)}^2 - 4 N_f z_{(11)^2} z_{(22)} + 4 N_f^2 z_{(11)^2}^2}{z_0^2}
    \\
    &\phantom{{}={}}
    - 3 \frac{\kappa^4 N_{\tau}}{2(N_c^2-1)N_c}
    \frac{8 N_f^2 z_{(11)^2} z_{(22)} - 2 N_f z_{(11)^2}^2 - 2 N_f z_{(22)}^2}{z_0^2} \\
    &\phantom{{}={}}
    - 3 \frac{\kappa^4 N_{\tau} (N_\tau -1)}{2 N_c^2}
    \frac{4 N_f^2 z_{(11)^2}^2 + 4 N_f^2 z_{(21)}^2 + 2 N_f z_{(21)} z_{(11)} + 
      2 N_f z_{(11)^2} z_{(21)}}{z_0^2} \\
    &\phantom{{}={}}
      + 30 \frac{\kappa^4 N_\tau}{N_c^2} 
      \frac{N_f z_{(11)}^2
        \left(
          2 N_\tau N_f z_{(11)^2} + (N_\tau-1) z_{(21)} +z_{(22)} + 2 z_{(11)} - 4 N_c
        \right) 
      }{z_{0}^3}\\
    &\phantom{{}={}}
    + 12 N_f \frac{\kappa^4 N_\tau}{N_c^2} \left(\frac{z_{(11)}}{z_{0}}\right)^3
    - 66 N_f^2 \frac{\kappa^4 N_\tau^2}{N_c^2} \left(\frac{z_{(11)}^4}{z_0^4}\right),
  \end{split}
\end{align}
where we have introduced the notation
\begin{align}
  z_{0} 
  & =
  \int\limits_{SU(N_c)} dW
  \det(1+h_1 W)^{2 N_f}\;, \\
  z_{(a_1 b_1) \ldots (a_k b_k)} 
  & =
  \int\limits_{SU(N_c)} dW
  \det(1+h_1 W)^{2 N_f} \prod_{i=1}^k \frac{(h_1 W)^{b_i}}{(1+h_1 W)^{a_i}}\;.
\end{align}
The required integrals are related in the following way,
\begin{align}
  z_{(11)} &= \frac{h_1}{2 N_f} \frac{\partial}{\partial h_1} z_{0}, \\
  z_{(22)} &= z_{(11)} - z_{(21)}, \\
  \label{eq:z-11-11-rel-z-11-z-21}
  z_{(11)^2} &= \frac{h_1}{2 N_f} \frac{\partial}{\partial h_1} z_{(11)} -
                \frac{1}{2 N_f} z_{(21)}\;.
\end{align}
Therefore, we only need to integrate
$z_{0}$, which corresponds to the integration over the static
determinant, and $z_{(21)}$.
Note that all integrands are class functions of $SU(N_c)$ group elements,
$f(W)=f(VWV^{-1})$, which are invariant under a change of basis. These functions only
depend on the eigenvalues $z_i$ of a group element. Furthermore, for our purposes
it is sufficient to specialise to functions which factorise in the following way
\begin{equation}
  f(W)= 
  \tilde{f}(z_1, \ldots, z_{N_c}) = 
  \sum_{\mu=1}^{N_c} 
  \tilde{f}_{1,\mu}(z_1) \cdot \ldots \cdot \tilde{f}_{N_c,\mu}(z_{N_c})\;.
\end{equation}
For such functions, the integration over the group can be expressed as 
\cite{Nishida:2003fb}
\begin{equation}
  \label{eq:facorized-haar-integral}
  \int\limits_{SU(N_c)} d{W} f(W) =
  \frac{1}{(2\pi)^{N_c}}
  \sum_{q=-\infty}^{\infty}
  \sum_{\mu=1}^{N_c}
  \det_{1\leq j,k\leq N_c}\left(\int\limits_{-\pi}^\pi d{\phi_j}\; 
  \tilde{f}_{j,\mu}(e^{i\phi_j}) e^{i(k-j+q)\phi_j}\right)\;.
\end{equation}%
Using this formula one obtains
\begin{align}
  \label{eq:z-0}
  z_{0}
  & =
  \sum_{p=0}^{2 N_f}
  \det_{1\leq i,j \leq N_c}\left(\binom{2 N_f}{i-j+p}\right)
  h_1^{p N_c}\;, \\
  \label{eq:z-21}
  z_{(21)}
  & = 
  \sum_{p=0}^{2 N_f} \sum_{\mu=1}^{N} \sum_{r=0}^{\infty}
  (-1)^r (r+1) \det_{1\leq i,j\leq N}\left(
    \begin{cases}
      \binom{2 N_f}{i-j+p-1-r} & \text{ if } i = \mu \\
      \binom{2 N_f}{i-j+p} & \text{ else}
    \end{cases}
  \right) h_1^{pN_c}\;.
\end{align}
To evaluate the occurring determinants we showed, using the techniques explained in
\cite{Krattenthaler1999}, that
\begin{equation}
  \label{eq:det-evaluated}
  \det_{1\leq i,j\leq N}\left(\binom{A}{L_i - j}\right) =
  (-1)^{\binom{N}{2}} \prod_{i=1}^N \frac{(A+N-i)^{\underline{L_i-i}}}{(L_i-1)!}
  \prod_{1\leq i<j\leq N} (L_i - L_j),
\end{equation}
where we have introduced the underline notation for the falling factorials
\begin{equation}
  n^{\underline{k}} = n \cdot (n-1) \cdots (n-k+1).
\end{equation}
Applying this formula results in
\begin{align}
  \label{eq:det-evaluated-special}
  \det_{1\leq i,j \leq N_c}\left(\binom{2 N_f}{i-j+p}\right)
  & = 
  \prod_{i=1}^p 
  \frac{(i-1+2 N_f-p+N_c)^{\underline{2 N_f-p}}}{(i-1+2 N_f-p)^{\underline{2 N_f-p}}}\;, \\
  \label{eq:z-21-evaluated}
  z_{(21)}
  & =
  \sum_{p=0}^{2 N_f}
  \det_{1\leq i,j \leq N_c}\left(\binom{2 N_f}{i-j+p}\right)
  \frac{p N_c (2N_f-p)(N_c+2N_f)}{2N_f(4N_f^2-1)} h_1^{p N_c}\;.
\end{align}

With the free energy density at hand, it is now possible to compute all thermodynamic functions 
for any desired value of $N_c$ within the framework of our hopping expansion.
Specifically we use the well known thermodynamic relations for the pressure
\begin{equation}
  a^4 p = -\frac{f}{N_{\tau}},
\end{equation}
baryon number density
\begin{align}
  a^3 n_B
  & =
  \frac{a^3}{N_c} \frac{T}{V} \frac{\partial\ln(Z)}{\partial\mu}\\
  & =
  - h_1 \frac{\partial f}{\partial h_1}
\end{align}
and energy density
\begin{align}
  a^4 e 
  & = 
  -\frac{1}{V} \left.\frac{\partial\ln(Z)}{\partial(1/T)}\right|_z \\
  & = 
  \frac{1}{N_{\tau}} \left(\frac{f}{h_1} \frac{\partial h_1}{\partial \kappa} + \frac{\partial f}{\partial \kappa} \right) a \frac{\partial \kappa}{\partial a}.
\end{align}
The derivative of $\kappa$ with respect to $a$ is computed at constant baryon
mass, which is given to first order in $\kappa$ by (\ref{eq:mlo}) resulting in
\begin{equation}
  a \frac{\partial \kappa}{\partial a} = \kappa \ln(2 \kappa).
\end{equation}

\subsection{The onset transition for increasing \texorpdfstring{$N_c$}{Nc}}

\begin{figure}[t]
\centering
\includegraphics[width=7.5cm]{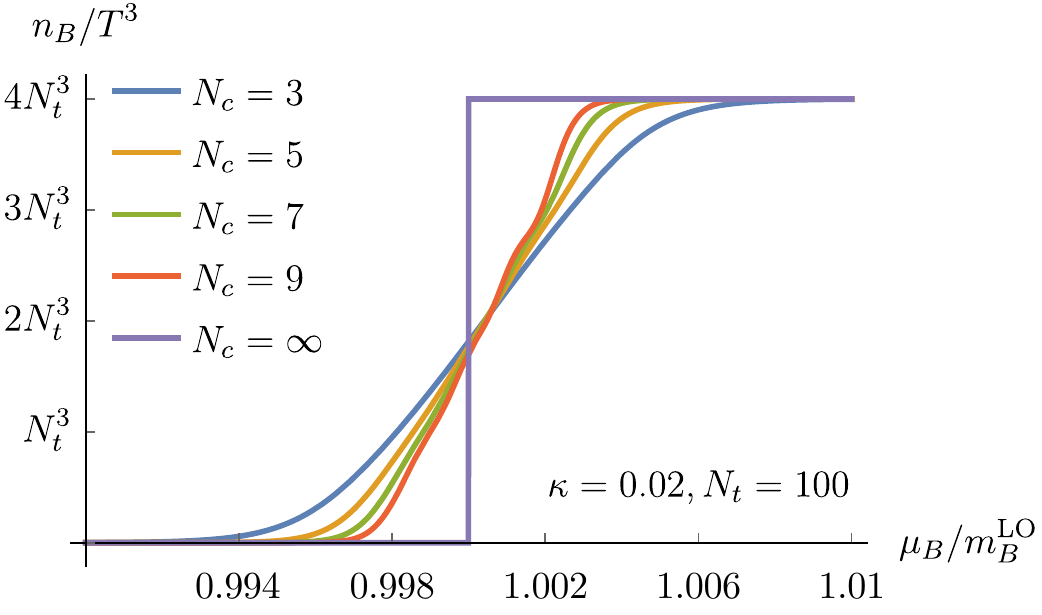}
\includegraphics[width=7cm]{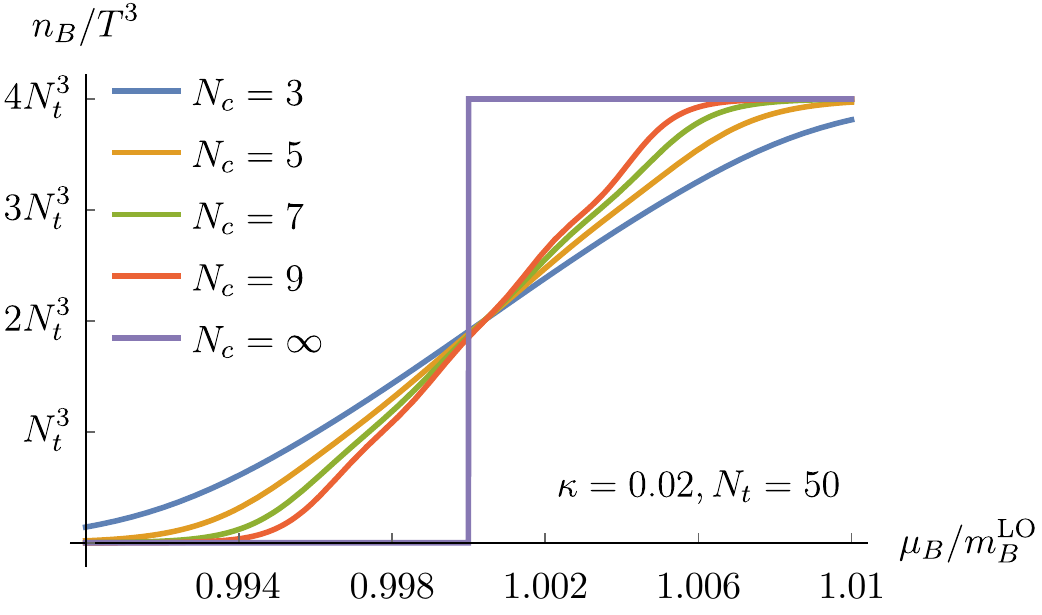}
\caption[]{Onset transition to baryon condensation for different values of $N_c$ and $N_\tau$ in the strong coupling limit.}
\label{fig:onset_nc_strong-coupling}       
\end{figure}
Of particular interest is the behaviour of the onset transition to finite baryon density,
which is shown in 
\fig\ref{fig:onset_nc_strong-coupling} for different choices of $N_c$.
We observe a steepening of the transition with increasing $N_c$, which asymptotically 
ends up in a step function, i.e.~a
first-order transition, even though we started with a smooth crossover at $N_c=3$. 

Decreasing the values of $N_\tau$, i.e.~increasing the temperature, 
flattens the curves with fixed $N_c$, but for asymptotically large $N_c$ a step
function is obtained for any finite starting value of $N_\tau$. 
Thus, growing $N_c$ appears to make the onset transition to baryon matter always first-order. 
(With increasing temperature one may question the neglect of $\lambda_1,\bar{h}_1$.
Their inclusion is discussed in sections \ref{sec:bh1}, \ref{sec:gauge}.)

Note that, in the limit of infinite $N_c$, the transition is between the vacuum and a saturated lattice,
similar to what happens in the static strong coupling limit at finite $N_c$. This saturated state is
a discretisation artefact and will move towards infinity as the continuum is approached, as we
discuss in section \ref{sec:cont}.

\subsection{Thermodynamic functions for large \texorpdfstring{$N_c$}{Nc}}

Since we have an explicit formula for the free energy for general $N_c$, one
can easily obtain the asymptotic behaviour of thermodynamic observables for
large $N_c$.  We study the behaviour of the different orders in the hopping
expansion separately.  This is necessary because, beyond the onset of baryon
condensation, the leading static term represents lattice saturation, which is
an unphysical artefact of discretisation. As discussed in section
\ref{sec:review}, correction terms do not contribute to saturation, but modify
the shape of the curves as they enter their low and high density asymptotes.
These effects will remain after continuum extrapolation and thus are physically
significant.

The general strategy for the asymptotic analysis is most easily illustrated for
the leading order contribution to the pressure at $N_f=1$
\begin{equation}
  a^4 p_{\mathrm{LO}} = \frac{1}{N_\tau} \ln\left(1 + (N_c+1)h_1^{N_c} + h_1^{2 N_c}\right).
\end{equation}
Note that, just like in the $SU(3)$ case in \eqref{eq:z-0-su3}, the prefactors
before $h_1^{N_c}$ can be understood from spin-degeneracy. Specifically, a
colour neutral state consisting of $N_c$ fermions is antisymmetric in colour
space under particle exchange. The only completely symmetric spin state of
$N_c$ spin $1/2$-particles is that with $s=N_c/2$ . States with this spin
and spin components $-N_c/2 \leq s_3 \leq N_c/2$ are degenerate, 
explaining the $N_c+1$ prefactor.

When $h_1<1$ then the term $h_1^{N_c}$ is strongly
suppressed (stronger than $N_c^k$ can grow for any $k$) and a Taylor expansion
around $h_1^{N_c}=0$ gives
\begin{align}
  a^4 p_{\mathrm{LO}} 
  & = 
  \frac{1}{N_\tau} (N_c+1)h_1^{N_c} + \mathcal{O}(h_1^{2 N_c}) \\
  & \sim
  \frac{1}{N_\tau} N_c h_1^{N_c} \text{ for } N_c \rightarrow \infty.
\end{align}
For $h_1>1$ the term with the highest power of $h_1^{N_c}$ determines
the asymptotic behaviour and one obtains
\begin{align}
  a^4 p_{\mathrm{LO}} 
  & \sim
  \frac{1}{N_\tau} \ln\left(h_1^{2 N_c}\right) \\
  & \sim
  \frac{2}{N_\tau} \ln(h_1) N_c.
\end{align}

For higher $N_f$ and higher orders the terms become more complicated, but the general
behaviour stays the same.  For $N_f=2$ our findings on both sides of the
onset transition are summarised in Table \ref{tab:scaling}. A clear picture
emerges: for $h_1<1$ all terms, due to the static determinant as well as the
corrections, come with a factor $h_1^{N_c}$ to some power. Since the fugacity
contains  $m_B^{\mathrm{LO}}/(N_c T)$  in the exponential, this factor will for
low temperatures always dominate the powers of $N_c$ and result in a stronger
exponential suppression  before the onset transition. In other words, the
curves for all quantities will be squeezed ever more tightly against the
chemical potential axis as $N_c$ gets large.
Since we do
not know the hopping expansion of the baryon mass  for general $N_c$, we
expressed our results in units of the leading order expression (\ref{eq:mlo}),
which is responsible for onset happening at $m_B^\mathrm{LO}=\mu_B$ at large $N_c$.

The more interesting situation is $h_1>1$, where we first focus on the baryon number density.  
As explained in Section \ref{sec:review}, the leading order
contribution in the hopping expansion corresponds to the static determinant only, 
for which the onset to baryon matter
is a first-order step function. This remains true for large $N_c$, 
with the lattice quark saturation density going as $a^3n^\mathrm{sat}=2N_fN_c$, 
i.e.~the baryon density behaves as $a^3n_B^\mathrm{sat}\sim \mathrm{const.}$. 

\begin{table}[tbp]
\centering
\begin{tabular}{|Sc|Sc||Sc Sc Sc|}
\hline
\multicolumn{2}{|c|}{Order hopping expansion} & $\kappa^0$ & $\kappa^2$ & $\kappa^4$ \\
\hline
\multirow{4}*{$h_1<1$} & 
$a^4p$ & $\sim \frac{1}{6 N_{\tau}} N_c^3 h_1^{N_c}$ & $\sim -\frac{1}{48} N_c^7 h_1^{2 N_c}$ & $\sim \frac{3 N_\tau \kappa^4}{800} N_c^8 h_1^{2 N_c}$ \\
    & $a^3n_B$ & $\sim \frac{1}{6} N_c^3 h_1^{N_c}$ & $\sim -\frac{N_{\tau}}{24} N_c^7 h_1^{2 N_c}$ & $\sim \frac{(9 N_\tau+1)N_\tau}{1200} N_c^8 h_1^{2 N_c}$ \\
    & $a^4e$ & $\sim -\frac{\ln(2\kappa)}{6} N_c^4 h_1^{N_c}$ & $\sim \frac{N_{\tau} \ln(2 \kappa)}{48} N_c^8 h_1^{2 N_c}$ & \\
    & $\epsilon$ & 0 & $\sim -\frac{1}{4} N_c^3 h_1^{N_c}$ & \\
\hline
\hline
\multirow{4}*{$h_1>1$} & 
 $a^4p$ & $\sim \frac{4 \ln(h_1)}{N_{\tau}} N_c$ & $\sim -12 N_c$ & $\sim 198 N_c$ \\
& $a^3n_B$ & $\sim 4 $ & $\sim - N_{\tau} \frac{N_c^4}{h_1^{N_c}}$ & $ \sim -\frac{(59 N_\tau - 19) N_\tau}{20} \frac{N_c^5}{h_1^{N_c}}$ \\
& $a^4e$ & $\sim -4 \ln(2 \kappa) N_c$ & $\sim 24 \ln(2 \kappa) N_c$ & \\
& $\epsilon$ & 0 & $\sim -6$  & \\
\hline
\end{tabular}
\caption[]{Large $N_c$ behaviour of the thermodynamic functions and the interaction energy
per baryon, order by order in the hopping expansion, 
on both sides of the onset transition for $N_f=2$.}
\label{tab:scaling}
\end{table}

The most intriguing result of this section is the $N_c$-scaling of the pressure
beyond baryon onset, $p\sim N_c$.  Preliminary results based on leading and
next-to-leading order were reported in \cite{scheunert}. Stability of this
finding through three orders suggests it to hold to any order in the hopping
expansion, and thus for all current quark masses. In this case strongly coupled
QCD beyond the onset transition to baryon matter satisfies the definition of
quarkyonic matter \cite{quarky}.  Note that there is a finite interaction energy
per baryon in units of baryon mass also for $N_c\rightarrow \infty$, as
was conjectured in \cite{quarky}. 
Its value at leading order $\kappa^2$ is determined  by
$d(d+1)/2$, where $d$ refers to the number of spatial dimensions. 

\subsection{The transition region \label{sec:trans}}

The results in the previous section were obtained by first fixing $h_1$,
i.e.~fixing the quark chemical potential, and then considering the limit 
$N_c\rightarrow \infty$. Right around the onset transition one can also consider 
$h_1^{N_c} \sim 1$. According to equation \eqref{eq:h1}, this means that the
quark chemical potential is adjusted such that $\mu-m \sim 1/N_c$, where
$am=\ln(2\kappa)$ is again the leading order constituent quark mass. In this
regime, the asymptotic behaviour is dominated by the prefactors of the
powers of $h_1^{N_c}$, which are polynomials in $N_c$, see equations
\eqref{eq:z-0}, \eqref{eq:z-21}, \eqref{eq:det-evaluated} and
\eqref{eq:z-21-evaluated}. For large $N_c$ one then obtains for the pressure in
the hopping expansion
\begin{equation}
  a^4p \sim 
  \frac{4}{N_\tau} \ln(N_c) - 
  3 \kappa^2 N_c + 
  \frac{(N_\tau-286)\kappa^4}{150} N_c^2 +
  \mathcal{O}(\kappa^6).
\end{equation}
This indicates that the hopping expansion does not converge for
large $N_c$. In \fig\ref{fig:onset_nc_strong-coupling} this shows up by the formation of uncontrolled wiggles in the
central region for sufficiently large $N_c$ (a first indication of this is seen for $N_c=9$).  
Thus, we cannot make a statement about the large $N_c$ 
behaviour in this window of parameter space. Fortunately, the width
of this region shrinks to zero for large $N_c$ and does not affect
the observations in the previous sections.
(For the same reason, 
it is not clear to us whether this behaviour
is related to a phase transition in $N_c$, as conjectured in \cite{lott}).

\subsection{Inclusion of \texorpdfstring{$\bar{h}_1$}{h1Bar}-corrections \label{sec:bh1}}

For statements at higher temperatures, or lower $N_\tau$, we also need to consider what
happens when $\bar{h}_1$ is included. In this case terms appear which mix $h_1$ and
$\bar{h}_1$. For $h_1>1$ the large $N_c$ analysis is clearly unchanged,
since in this case the highest powers of $h_1^{N_c}$ determine the asymptotic
behaviour. For $h_1<1$ $\mu$-independent terms with equal power of $h_1$ and
$\bar{h}_1$ become relevant as they are not suppressed when $N_c \rightarrow
\infty$. However, these contributions are of mesonic nature and expected
to scale as $\sim N_c^0$.

This can be shown explicitly for $N_f=1$ at leading order in the hopping
expansion, as the contribution of the static determinant is known to be
\cite{Unger:2014oga}
\begin{align}
  \begin{split}
    -f_{h_1,\bar{h}_1} 
    & =
    \ln\biggl(\;\sum_{k=0}^{2 N_c} T(k) (2 \kappa)^{2 k N_\tau} \\
    & \qquad +
    \sum_{k=0}^{N_c} P(k) (2 \kappa)^{(2 k + N_c) N_\tau} 2 \cosh(N_c \mu/T) +
    (2\kappa)^{2 N_c N_\tau} 2 \cosh(2 N_c \mu/T)\biggr)\;,
  \end{split}
\end{align}
with $T(k) = \binom{\min(k,2 N_c-k)+3}{3}$ and $P(k)=(N_c+1-k)(k+1)$. For
$\mu<a \ln(2\kappa)$ (i.\,e.~$h_1<1$) the $\mu$-dependent terms vanish as
$N_c\rightarrow\infty$. Obviously, this is not the case for the
$\mu$-independent terms. Their contribution can be evaluated for $N_c
\rightarrow \infty$ by using the following variant of the geometric series
(which can be obtained by differentiation):
\begin{equation}
  \sum_{k=0}^\infty (k+l)^{\underline{l}} x^k = \frac{l!}{(1-x)^l} \text{ for } \lvert x \rvert <1.
\end{equation}
This results for $N_c\rightarrow\infty$ in the pressure
\begin{equation}
  a^4 p_{h_1,\bar{h}_1} = -\frac{4}{N_\tau} \ln\left(1-(2\kappa)^{2 N_\tau}\right) \sim N_c^0.
\end{equation}

\subsection{Gauge corrections and 'tHooft scaling \label{sec:gauge}}

Our discussion so far has been for the strong coupling limit $\beta=0$, and so
does {\it not} correspond to the intended  't Hooft limit of large $N_c$. 
In this section we
argue that our observations carry over to the t'Hooft limit once we include
gauge corrections, at least to the orders considered.

Including the leading gauge corrections to the fermion determinant, as well as partial resummation of the 
corresponding diagrams to
all orders, proceeds in the same manner as in \cite{bind} for any $N_c$.
Through $\mathcal{O}(\kappa^2)$ the free energy density now reads 
\begin{equation}
  -f = \ln(z_0(h_1)) + 
    \frac{\kappa^2 N_\tau}{N_c}\left[1+2 \frac{u-u^{N_\tau}}{1-u}\right](-6 N_f)
    \frac{z_{11}(h_1)}{z_0(h_1)}^2,
\end{equation}
where $u$ can be computed using (\ref{eq:character-coeff-all-nc}) and $h_1$ now includes
corrections,
\begin{equation}
  h_1 = 
  (2 \kappa)^{N_\tau} e^{a \mu N_\tau}
  \exp\left[6 N_\tau \kappa^2 \frac{u-u^{N_\tau}}{1-u}\right].
\end{equation}
In taking the 't Hooft limit, i.\,e.~keeping $\lambda_H = 2 N_c^2/\beta$
fixed, one has for $\lambda_H>1$ \cite{gw,Drouffe:1983fv}
\begin{equation}
  u(\beta) = \frac{1}{\lambda_H}\;.
\end{equation} 
Therefore, these gauge corrections only modify the asymptotic behaviour of the
thermodynamic functions by a constant $\sim N_c^0$.

Furthermore, we also consider the leading order contribution from the pure
gauge sector to the effective theory (the first line in equation~\eqref{zpt})
with 
\begin{equation}
  \lambda_1 = u^{N_\tau}\;,
\end{equation}
to leading order in the character expansion. The first correction to $-f$ in equation
(\ref{eq:free-energy}) due to the inclusion of this term, combined with the centre-symmetric part of 
the static determinant, reads
\begin{equation}
  -f_{\lambda_1,h_1} = 6 \lambda_1 \frac{z_{(01)} z_{(0\;-1)}}{z_0^2}.
\end{equation}
Employing the previously outlined
integration techniques one obtains
\begin{align}
  z_{(01)}
  & = 
  \sum_{p=0}^{2 N_f}
  \det_{1\leq i,j \leq N_c}\left(\binom{2 N_f}{i-j+p}\right)
  \frac{p N_c}{N_c+2 N_f-p} h_1^{p N_c}\;, \\
  z_{(0\;-1)}
  & = 
  \sum_{p=0}^{2 N_f}
  \det_{1\leq i,j \leq N_c}\left(\binom{2 N_f}{i-j+p}\right)
  \frac{(2 N_f-p)N_c}{N_c+p} h_1^{p N_c}\;.
\end{align}
In the 't Hooft limit $\lambda_1=\frac{1}{\lambda_H^{N_t}}$, and
therefore the asymptotic analysis of this term can be done in the same way as for the
strong coupling contributions. The result is
\begin{equation}
  a^4 p_{\lambda_1,h_1} \sim \frac{4}{N_\tau \lambda_H^{N_\tau}} N_c^3
  \begin{cases}
    h_1^{N_c}, &\text{ if } h_1<1\;, \\
    \frac{1}{h_1^{N_c}}, &\text{ if } h_1>1\;.
  \end{cases}
\end{equation}
Hence, the $N_c$ scaling of these corrections is subleading for $h_1>1$, while
for $h_1<1$ the previous results are again only modified by a constant $\sim
N_c^0$. 
Starting at $\mathcal{O}(\lambda_1^4)$ there are contributions which are entirely due to the pure gauge
part of the action. For these contributions only integrals of the type
\begin{equation}
  \int\limits_{SU(N_c)} dW \tr(W)^n \tr(W^\dagger)^n = n! \quad \text{ for }\quad n \leq N_c
\end{equation}
are relevant. When taking the large $N_c$ limit, order by order these
contributions to the pressure
are $\mu$-independent and 
scale as $\sim \lambda_1^k \sim N_c^0$.

The last statement hinges on the fact that the $N_c$ dependence of
$\lambda_1$ is solely determined by $u$.  In \cite{myers} corrections to
$\lambda_1$ to $\mathcal{O}(u^8)$ were computed and, although some corrections
do introduce additional $N_c$ factors, those cancel order by order when all
corrections are summed up. A similar observation, including higher
representations, was made in \cite{p_nc} in the context of a strong
coupling expansion of pure gauge theory without using an effective theory.

\begin{figure}[t]
\centering
\includegraphics[width=7cm]{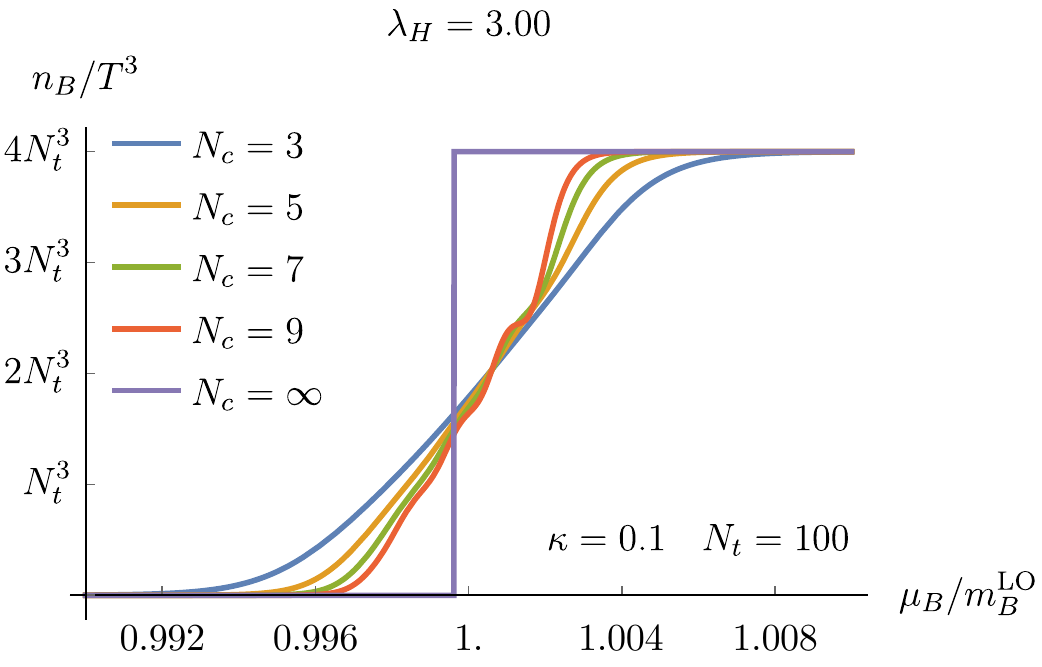}
\includegraphics[width=7cm]{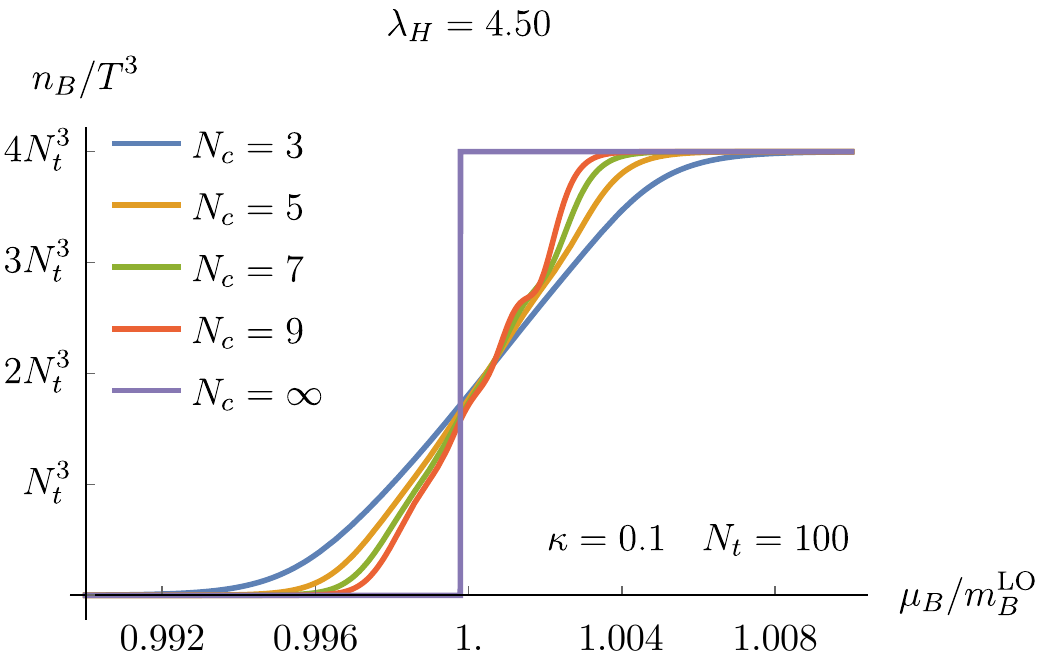}
\caption[]{Baryon density, including gauge corrections, for growing $N_c$ with the 't Hooft coupling held fixed. The qualitative
behaviour is the same as in the strong coupling limit.}
\label{fig:hooft}       
\end{figure}

The influence of the gauge corrections is illustrated in \fig\ref{fig:hooft}
for two different choices of the 't Hooft coupling.  Clearly, the small quantitative
modifications by the gauge corrections do not alter the qualitative
$N_c$-behaviour observed earlier.  Of course, in higher orders the situation
might be more complicated, as new interactions can arise in the effective
theory with non-trivial $N_c$-dependence.  Nevertheless, the dominant
contribution to the large $N_c$ limit of baryon dynamics should always be
represented by powers of $h^{N_c}$ like in the leading contributions considered
here, since $N_c$ occurs in the exponent in these cases. 

\subsection{Approaching the continuum \label{sec:cont}}

\begin{figure}[t]
\centering
\begin{subfigure}[]{\textwidth}
\includegraphics[width=7cm]{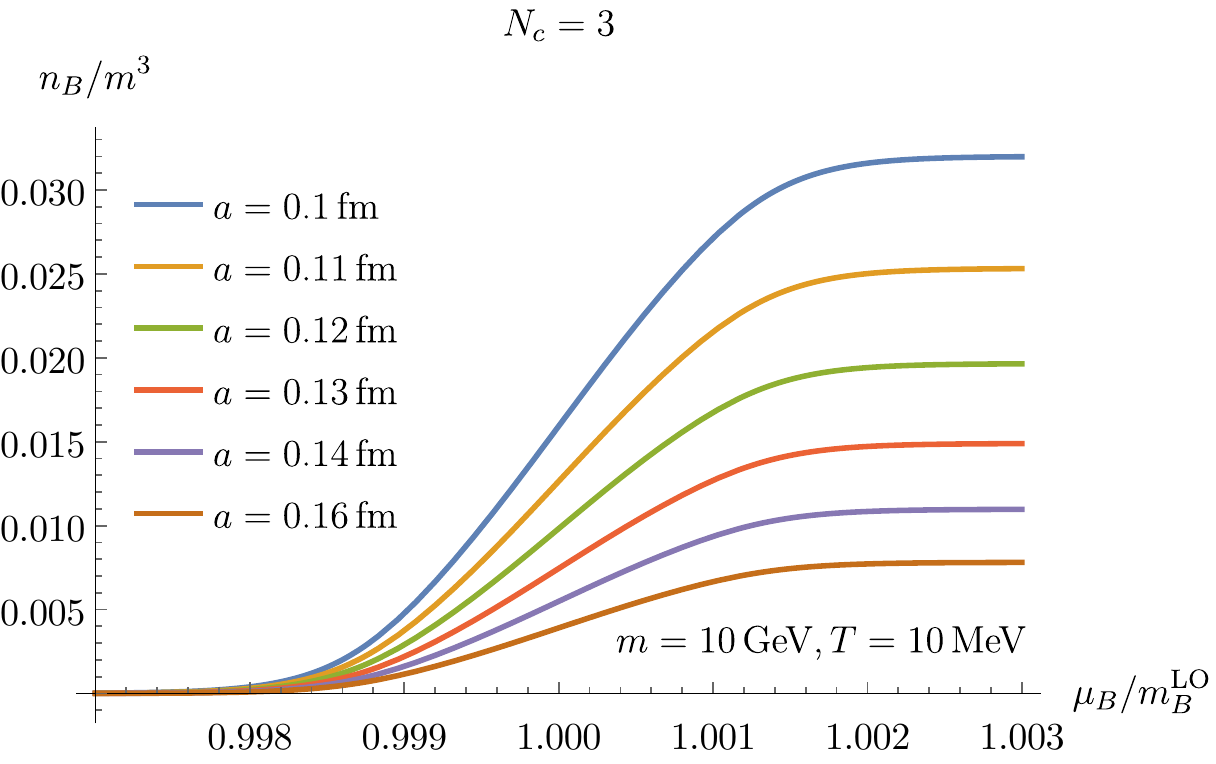}
\includegraphics[width=7cm]{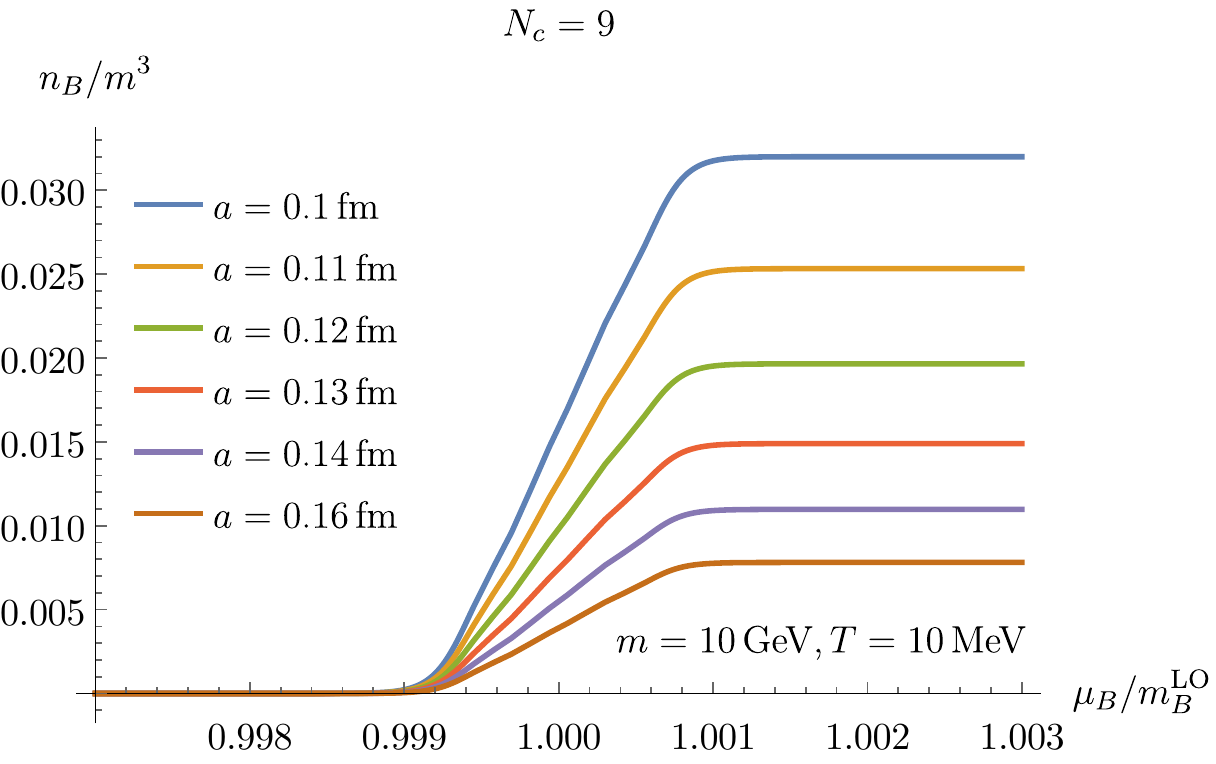}
\\\\
\includegraphics[width=7cm]{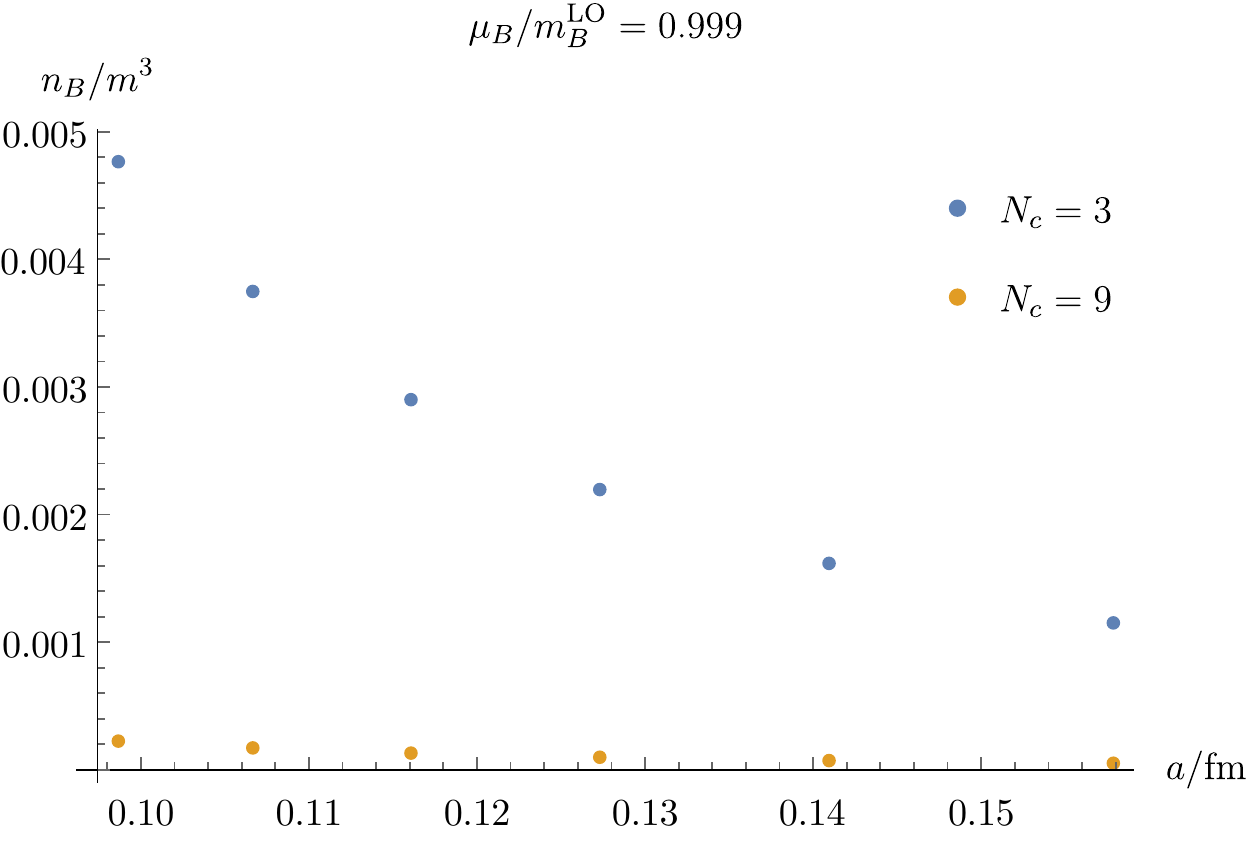}
\includegraphics[width=7cm]{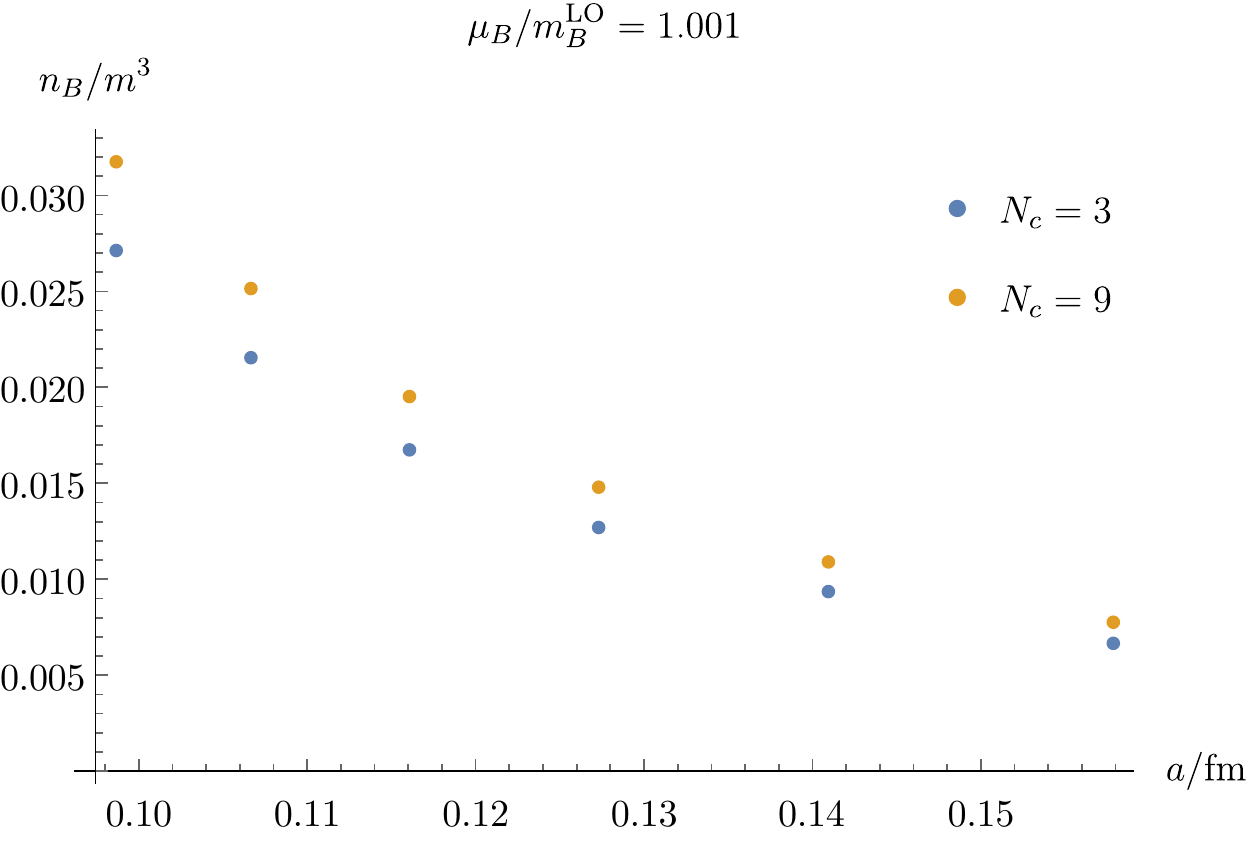}
\end{subfigure}
\caption[]{Continuum approach of the onset transition: steepening of the transition with growing $N_c$ is
also obtained if the continuum limit is taken first.}
\label{fig:cont}       
\end{figure}
This situation leaves, however, one caveat.  Even if one would be able
to include gauge contributions to all orders, the interchange of the
$N_c\rightarrow\infty$ limit and the strong coupling expansion was observed to
be ``highly suspicious'' in the case of QCD in 1+1 dimensions \cite{gw}.  Our
analysis so far was based on taking $N_c$ large before a continuum limit.  The
fact that the density at the onset transition immediately jumps to lattice
saturation indeed suggests that the limits should be taken in the opposite
order, if one is interested in continuum physics.  In this case, the interplay
between large $N_c$ and the Pauli principle should lead to a finite continuum
density, just as for $N_c=3$. 

To get an idea if our results are consistent with this expectation we
investigated the behaviour of the baryon density towards the continuum. To set
the scale at $SU(3)$ we use the same strategy as in \cite{bind}, which gives a
rough estimate of the parameter space. At first, since heavy quarks have little
influence on the running of the coupling, we use the non-perturbative
beta-function of pure gauge theory to get a relation between $\beta_{SU(3)}$
and $a/r_0$, where $r_0$ is the Sommer parameter \cite{Necco:2001xg}. Using
$r_0=0.5$ fm sets a physical scale for the lattices and the temperature can be
adjusted by $N_\tau$ via $T=1/(aN_\tau)$. To obtain the corresponding $\beta$
for $SU(N_c)$ we keep $\lambda_H = 2 N_c^2/\beta_{SU(N_c)} = 18/\beta_{SU(3)}$
fixed. Finally, the leading order expression \eqref{eq:mlo} is used to keep the
constituent quark mass constant for different $a$.

\begin{figure}[t]
\centering
\begin{subfigure}[]{\textwidth}
\includegraphics[width=7cm]{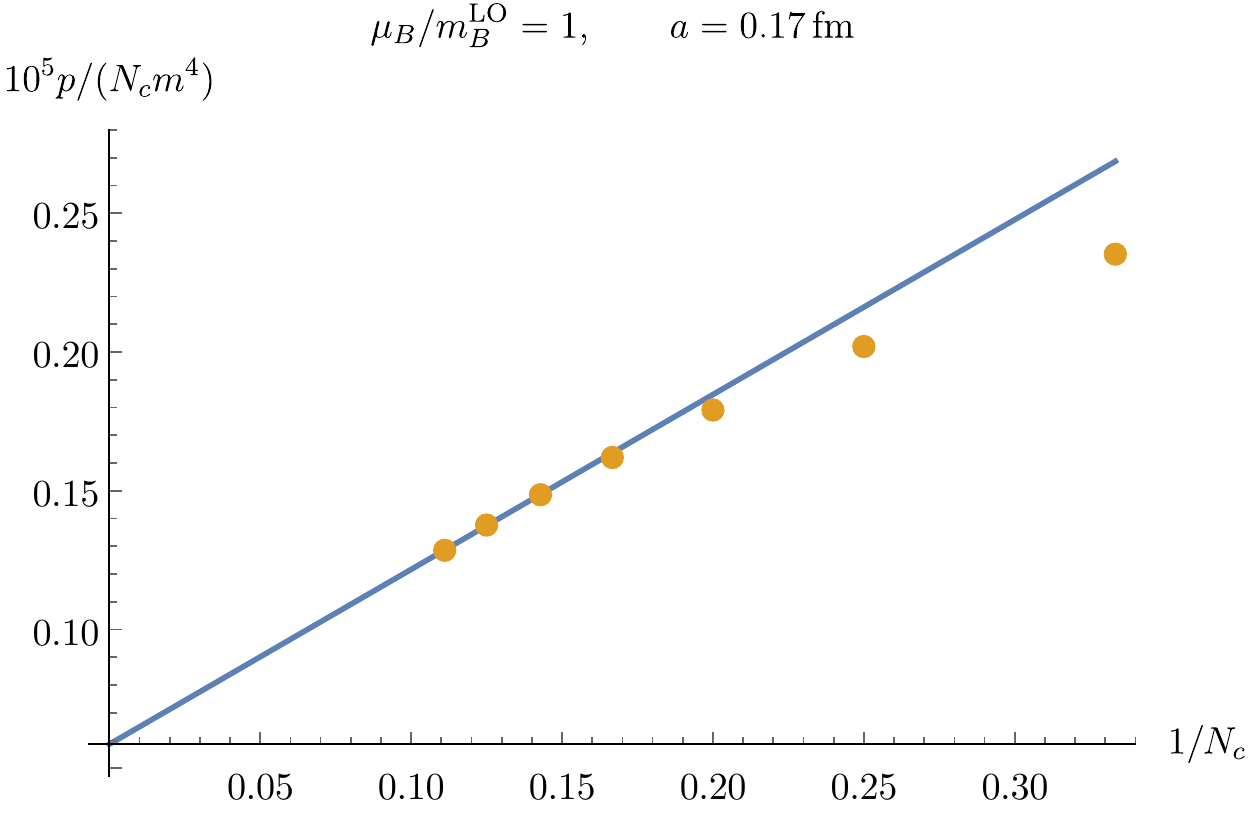}
\includegraphics[width=7cm]{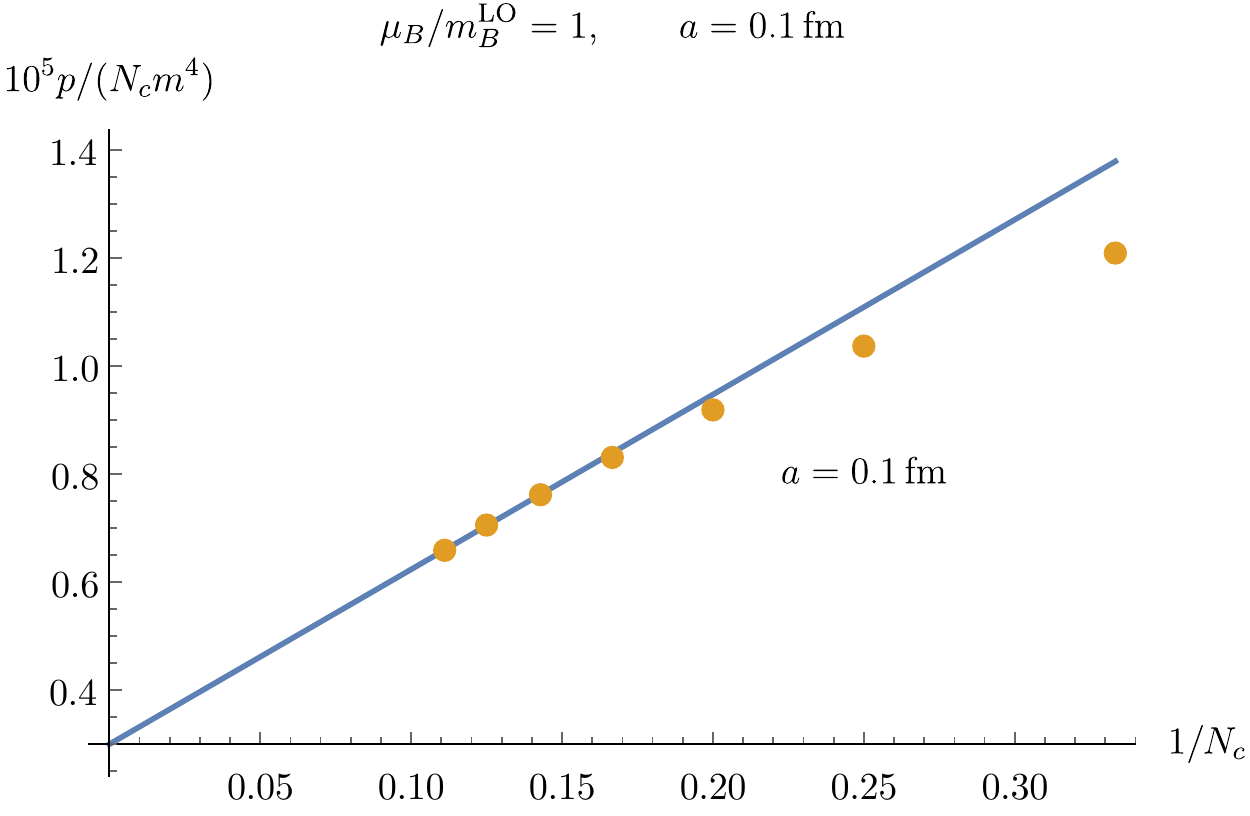}
\end{subfigure}
	\caption[]{The $N_c$-scaling of the pressure, $p\sim N_c(1+{\rm const.\;}N_c^{-1}+\ldots)$,
	for two different lattice spacings in the interval $N_c\in[3,9]$.}
\label{fig:pcont}       
\end{figure}

The outcome of this procedure is illustrated in \fig\ref{fig:cont}, where the
lattice spacing is varied for fixed $N_c=3,9$. 
Each continuum extrapolation leads to a finite value for the density, which is
smaller or larger, for $h_1<1$, $h_1>1$, respectively, as $N_c$ is increased. 
It is thus apparent that the transition steepens with growing $N_c$ also 
when the continuum is approached first. Similarly, 
the pressure is shown 
in \fig\ref{fig:pcont} for two different lattice spacings and $N_c\in[3,9]$.
Since this is far from the large $N_c$ limit, we
explicitly checked that the first subleading contribution goes 
as $\sim N_c^0$. The figure shows that the full result follows the predicted $N_c$-scaling to hold
in a region where the lattice is only about half filled, i.e., not yet dominated
by lattice saturation.
This behaviour appears to be stable
as the lattice is made finer.
While for $N_c=3$ and sufficiently heavy quarks
a continuum limit can be explicitly taken 
\cite{bind,k8}, for large $N_c$ this is difficult in practice, because in 
the problematic transition region, cf.~section \ref{sec:trans}, 
the required length of the hopping series grows exponentially with $N_c$.
We are therefore unable to
explicitly demonstrate first-order behaviour or the scaling 
$p\sim N_c$ of the large $N_c$-limit in the
continuum.

\subsection{The phase diagram with growing \texorpdfstring{$N_c$}{Nc}}

We have seen that, in the strong coupling limit, the onset transition to baryon matter becomes first-order
for any temperature.
In the last section we provided evidence for a steepening of the liquid gas transition with $N_c$ also in
the continuum. While we are unable to take the continuum and large $N_c$ limits in this order, we 
now argue on physical grounds that the onset transition has to become first order when $N_c\rightarrow \infty$.
Within the effective theory (as well as physical QCD), 
the endpoint of the nuclear liquid gas transition is located  where the temperature starts to exceed the 
binding energy per baryon. In table \ref{tab:scaling}, we found the interaction energy in units of baryon mass
to scale as $\epsilon\sim \mathrm{const}$. Consequently, the binding energy in $N_c$-independent units  
scales as $\sim N_c$
as expected, and the critical endpoint of the liquid gas transition moves to ever larger temperatures. 
On the other hand, the deconfinement transition temperature $T_d$ is only sensitive to meson physics, and 
in the large $N_c$-limit is within $\sim N_c^{-2}$ of its value 
at $N_c=3$ \cite{teper}. Hence, in the limit $N_c\rightarrow \infty$, 
temperatures in the range $0<T<T_d$ never exceed the binding energy between baryons and the onset 
transition must be of first order. 

\begin{figure}[t]
\centering
\includegraphics[width=7cm]{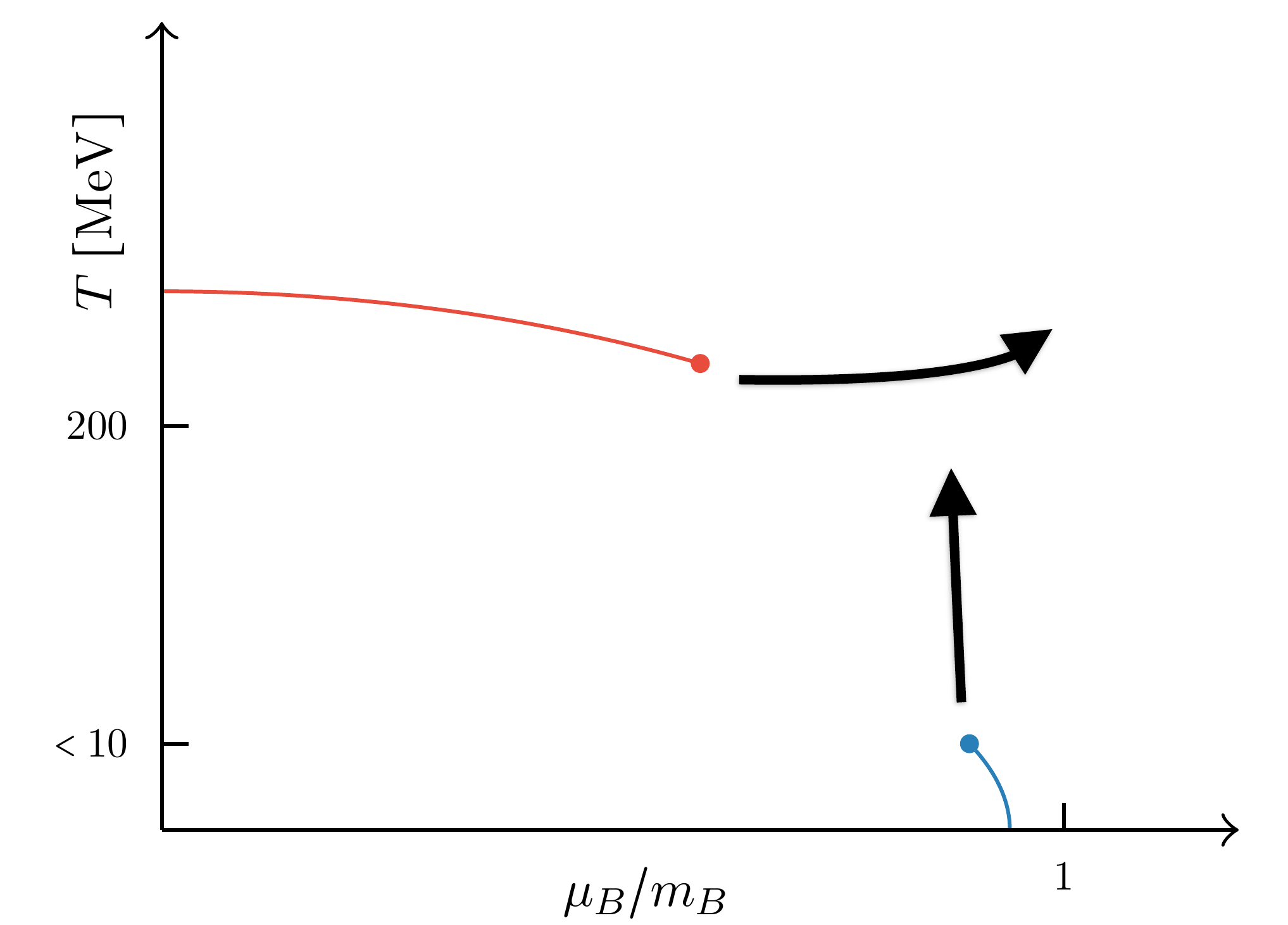}
\caption[]{ Arrows indicate the smooth change of the phase transition lines
with growing $N_c$.}
\label{fig:pd_move}       
\end{figure}

This implies that the critical
endpoint of the baryon onset transition increases with growing $N_c$ until it hits 
another discontinuity, 
as indicated in \fig\ref{fig:onset_nc_strong-coupling} (right).
We know already from perturbation theory 
that also the deconfinement transition line ``straightens out'' with growing $N_c$,
as the deconfinement transition becomes less and less sensitive to the quark contributions.
Altogether, we then observe how the predicted rectangular
phase diagram of \fig\ref{fig:pd_nc} emerges continuously by increasing $N_c$, as indicated in 
\fig\ref{fig:pd_move} .

\subsection{Quarkyonic matter on the lattice?}

While lattice saturation is a mere discretisation artefact 
and may seem uninteresting from a continuum perspective, it adds an intriguing feature here. 
The lattice saturation density is clearly
determined by the quark degrees of freedom. 
Besides counting degrees of freedom, this follows also from the fact that the saturation 
density is precisely the same in the case of large isospin chemical potential \cite{bind}.
Thus, approaching saturation, the 
lattice is filled with quark matter. 

On the other hand, the onset transitions at finite baryon as well as  
isospin chemical potentials are related to the condensation of hadrons, and not quarks. 
This follows from the fact that
the critical chemical potential is different in the baryonic and isospin cases \cite{bind}, i.e.~$m_B/3\neq m_\pi/2$.

For increasing baryon chemical potential, a lattice filling up with baryon number 
is thus consistent with
the picture of quarkyonic matter, in the sense that it shows a smooth transition
from baryon matter to quark matter in a remarkably narrow range of chemical potentials.
As the lattice is made finer, the saturation level in physical units increases and  is 
reached at larger chemical potentials. Eventually, in the continuum limit, the interplay between
the attractive baryon interaction and Pauli repulsion will
lead to the physical saturation density known from nuclear matter, 
while
the quark matter observed at lattice saturation gets shifted to larger chemical 
potentials (possibly infinite), with a quarkyonic regime in between.

\section{Conclusions  \label{sec:phys}}

We have studied the large $N_c$-behaviour of QCD in the cold and dense regime within an effective
lattice theory derived by combined strong coupling and hopping expansions, 
which is valid for sufficiently heavy quarks. At low temperatures
and $\mu_B\sim m_B$ it exhibits a transition to baryon condensation, which is the heavy quark
analogue of the nuclear liquid gas transition. 
By considering the effective theory for general gauge group $SU(N_c)$ we have shown that,
in the strong coupling limit and through three consecutive orders 
in the hopping expansion, 
the pressure
in the baryon condensed phase scales as $p\sim N_c$ and the onset transition
becomes first-order for large $N_c$. This behaviour is
stable under inclusion of the leading gauge corrections. We have pointed out that the
continuum limit has to be taken {\it before} the large $N_c$ limit, which makes definite
conclusions for continuum physics much harder to reach. Nevertheless, we have shown 
our findings to be stable also in this ordering for the range of lattice spacings and 
$N_c$-values we were able to consider with our truncated series. Our results are
thus consistent with the large $N_c$ phase diagram and the definition of quarkyonic
matter proposed in \cite{quarky}.

For $N_c=3$, onset to baryon matter happens at $\mu_B^c<m_B$ (because of the binding energy
between baryons) whereas $h_1=1$ at some $\mu_B>m_B$. 
The onset transition then marks the condensation of baryons, which smoothly
turn into
quarkyonic matter, whose pressure scales as $p\sim N_c$ and whose effective degrees
of freedom can in principle change smoothly from baryon-like to quark-like as a function 
of chemical potential.
 
The QCD parameter values realised by nature are $m_{u,d}\sim 2-5$ MeV,
in contrast to the heavy quarks on which our analysis above is based. 
What can be
said about the physical situation?
As remarked in section \ref{sec:nc}, the large $N_c$ analysis is independent
of the current quark masses, with $m_B\sim N_c$ always and Feynman diagrams with quark
loops suppressed. Thus, whether there is a deconfinement transition, 
a chiral transition or a crossover
at $N_c=3$ is immaterial for the forming of a first-order horizontal deconfinement 
line at large $N_c$. 
On the other hand, the baryon onset transition 
is also present in a different effective lattice theory derived from QCD, 
which is valid in the chiral limit and the strong coupling regime \cite{unger}. 
(Moreover, it is also expected from nuclear physics \cite{kapusta}.) 
If our results generalise to all orders in the hopping expansion 
and are stable 
in the proper order of limits, 
then we expect
the $N_c$-scaling of the pressure
as well as the evolution of the baryon onset transition with growing $N_c$ to look
qualitatively the same when starting
from the physical situation, 
i.e. these would be genuine features of $SU(N_c)$-QCD. 
Whether or not there is also a chiral transition 
at some larger chemical potential cannot be decided within the 
current framework, 
but requires additional 
investigations in an effective theory including chiral symmetry, such as \cite{unger}.

\acknowledgments We thank J.Glesaaen for collaboration during the initial stages of this work, and 
M.~Alford and A.~Schmitt for enlightening discussions.
The authors acknowledge support by the Deutsche Forschungsgemeinschaft (DFG) through the grant CRC-TR 211 ``Strong-interaction matter
under extreme conditions'' and by the Helmholtz International Center for FAIR within the LOEWE program of the State of Hesse.

\end{document}